\documentclass[12pt,preprint]{aastex}
 
\shorttitle{Off-Axis Jet Model for XRF}
\shortauthors{Yamazaki et al.}
 
\newcommand{\beq}{\begin{equation}}
\newcommand{\beqa}{\begin{eqnarray}}
\newcommand{\eeq}{\end{equation}}
\newcommand{\eeqa}{\end{eqnarray}}

\def\Ep{E_{\rm p}}
\def\Eiso{E_{\rm iso}}
\def\Eg{E_\gamma}

\def\max{{\rm max}}
\def\min{{\rm min}}
\def\intrin{{|_{z=0}^{\theta_v=0}}}
\def\malpha{{\langle\alpha\rangle}}

\begin{document}

\title{Peak Energy--Isotropic Energy Relation
in the Off-Axis Gamma-Ray Burst Model}


\author{Ryo~Yamazaki\altaffilmark{1},
Kunihito~Ioka\altaffilmark{2},
and Takashi~Nakamura\altaffilmark{1}}
\altaffiltext{1}{Department of Physics,
Kyoto University, Kyoto 606-8502, Japan}
\altaffiltext{2}{Department of Earth and Space Science,
Osaka University Toyonaka 560-0043, Japan}
\email{
yamazaki@tap.scphys.kyoto-u.ac.jp,
ioka@vega.ess.sci.osaka-u.ac.jp,
takashi@tap.scphys.kyoto-u.ac.jp
}

\def\E{{\cal E}}
\def\d{{\rm d}}
\def\p{\partial}
\def\w{\wedge}
\def\o{\otimes}
\def\f{\frac}
\def\tr{{\rm tr}}
\def\Half{\frac{1}{2}}
\def\half{{\scriptstyle \frac{1}{2}}}
\def\T{\tilde}
\def\RA{\rightarrow}
\def\N{\nonumber}
\def\n{\nabla}
\def\bb{\bibitem}
\def\BE{\begin{equation}}
\def\EE{\end{equation}}
\def\BEA{\begin{eqnarray}}
\def\EEA{\end{eqnarray}}
\def\L{\label}
\def\VVM{\langle V/V_{\rm max}\rangle}
\def\zero{{\scriptscriptstyle 0}}
\begin{abstract}
Using a simple uniform jet model of
prompt emissions of gamma-ray bursts (GRBs),
we reproduce the observed peak energy--isotropic energy relation.
A Monte Carlo simulation shows that
the low isotropic energy part of the relation is
dominated by events viewed from off-axis directions,
and the number of the off-axis events is
about one-third of the on-axis emissions.
We also compute the observed event rates of the GRBs,
the X-ray-rich GRBs, and the X-ray flashes detected by {\it HETE-2},
and we find that they are similar.
 
\end{abstract}
 
 
%
\section{Introduction}
 
There is a strong correlation between the rest-frame
spectral peak energy $(1+z)\Ep$ and the isotropic 
equivalent $\gamma$-ray energy $\Eiso$ of the 
gamma-ray bursts (GRBs).
This relation ($\Ep$--$\Eiso$ relation) was first discovered
by \citet{amati2002}, and recently extended down to lower
energies \citep{atteia2003,lamb2003,sakamoto2003},
so that  $\Eiso$ ranges over 5 orders of magnitude.
A similar relation, the $\Ep$--luminosity relation is also found by
\citet{yonetoku2003a}, and both relations could become a new distance
indicator.
The geometrically corrected
$\gamma$-ray energies $E_\gamma=(1-\cos\Delta\theta)\Eiso$
narrowly cluster around a standard energy $E_\gamma\sim10^{51}$~ergs
\citep{bloom2003ser,frail2001}, so that the opening half-angle of the jet
in the on-axis uniform jet model ranges 2.5 orders of magnitude.
This  means that if the low isotropic energy events correspond to the
wide opening half-angle jet,
the jet opening half-angle of the typical GRBs $\Delta\theta$ becomes
less than $1^{\circ}$ \citep{lamb2003xrf}.
However such a small angle jet has difficulties in the 
standard afterglow models and observations.
 \citep[see also][]{zhang2003st}.
 
The low-energy (low-$\Ep$) part of the relation consists of
X-ray flashes (XRFs), that were identified by
{\it BeppoSAX} \citep{heise2001} and other satellites
\citep{stro98,gotthelf1996,hamilton1996,arefiev2003} and
have been accumulated by {\it HETE-2} \citep{barraud2003}.
Theoretical models of the XRF have been widely
discussed \citep{yamazaki2003cosmo}:
{\it high-redshift GRBs} \citep{heise2001,barraud2003},
{\it wide opening angle jets} \citep{lamb2003xrf},
{\it internal shocks with small contrast of high Lorentz factors} 
\citep{mochkovitch2003,daigne2003},
{\it failed GRBs or dirty fireballs} 
\citep{dermer1999,huang2002,dermer2003},
{\it photosphere-dominated fireballs}
\citep{meszaros2002,ramirez2002,drenkhahn2002},
{\it peripheral emissions from collapsar jets}
\citep{zhang2003collaps}
and {\it off-axis cannonballs} \citep{dar2003}.
The issue is what is the main population among them.

We have already proposed {\it the off-axis jet model}
\citep{yamazaki2002,yamazaki2003cosmo}.
The viewing angle is the key parameter to understanding the
various properties of the GRBs and may cause various relations
such as the luminosity-variability/lag relation,
the $\Ep$-luminosity relation and the luminosity-width relation
\citep{ioka2001,salmonson2002}.
When the jet is observed from off-axis, it looks like
an XRF because of the weaker blue-shift than the GRB.

There are some criticisms against our off-axis jet model.
The original version of our model \citep{yamazaki2002} required the
source redshift to be less than $\sim 0.4$ to be bright enough for
detection, conflicting with the observational implications
\citep[e.g.,][]{heise2002talk,bloom2003xrf}.
\citet{yamazaki2003cosmo} showed that higher redshifts ($z\gtrsim1$)
are possible with narrowly collimated jets ($\lesssim0.03$~rad), while
such small jets have not yet been inferred by afterglow observations
\citep{bloom2003ser,panaitescu2002,frail2001}.
The luminosity distance to the sources at $z\sim0.4$
is $d_L\sim2$~Gpc, which is only a factor of 3 smaller than that at
$z\sim1$ (corresponding to $d_L\sim7$~Gpc).
Thus small changes of  parameters in our model allow us to
extend the maximum redshift of the off-axis jets to $z\gtrsim1$
even for $\Delta\theta\sim0.1$.
This will be explicitly shown in \S~\ref{sec:index}.
Therefore, off-axis events may represent a large portion
of whole observed GRBs and XRFs since the solid angle to which the off-axis events
are observed is large.
 
In this Letter, taking into account the viewing angle effects,
we derive the observed $\Ep$--$\Eiso$ relation in our
simple uniform jet model.
This paper is organized as follows.
In \S~\ref{sec:model}
we describe a simple off-axis jet model for the XRFs.
Then, in \S~\ref{sec:index}, it is shown that
the off-axis emission from the cosmological sources
can be observed, and
the $\Ep$--$\Eiso$ relation is discussed in
\S~\ref{sec:EpEiso_XRF}.
Section~\ref{sec:dis} is devoted to discussions.
We also show that the observed event rates of GRBs and XRFs
are reproduced in our model.
Throughout the Letter,  we adopt a $\Lambda$CDM cosmology
with $(\Omega_{m}, \Omega_{\Lambda}, h)
=(0.3, 0.7, 0.7)$.

\section{Prompt emission model of GRBs}
\label{sec:model}
We use a simple jet model of prompt emission of GRBs
considered in our previous works 
\citep{yamazaki2003cosmo,yamazaki2003low}. 
We assume a uniform jet with a sharp edge, whose properties do not
vary with angle.
Note that the cosmological effect is included in these works
\citep[see also][]{yamazaki2002,yamazaki2003df,ioka2001}.
We adopt an instantaneous emission, at $t=t_0$ and $r=r_0$,
of an infinitesimally thin shell moving with the Lorentz factor
$\gamma$.
Then the observed flux of a single pulse at frequency $\nu=\nu_{z}/(1+z)$
and time $T$ is given by
\begin{eqnarray}
F_{\nu}(T)
=\f{2(1+z)r_0 c A_0}{d_L^2}
{{\Delta \phi(T) f\left[\nu_z\gamma (1-\beta\cos\theta(T))\right]
}\over{\left[\gamma (1-\beta\cos\theta(T))\right]^2}},
\label{eq:jetthin}
\end{eqnarray}
where $1-\beta\cos\theta(T)=(1+z)^{-1}({c\beta}/{r_0})(T-T_0)$
and $A_0$ determines the normalization of the emissivity.
The detailed derivation of equation (\ref{eq:jetthin}) and the
definition of $\Delta\phi(T)$
are found in Yamazaki et al. (2003b).
In order to have a spectral shape similar to
the observed one \citep{band1993},
we adopt the following form of the spectrum in the comoving frame,
\begin{equation}
f(\nu')= \left\{
\begin{array}{ll}
(\nu'/\nu'_0)^{1+\alpha_B}\exp(-\nu'/\nu'_0)
 & {\rm for} \ \nu'/\nu'_0 \le \alpha_B - \beta_B \\
(\nu'/\nu'_0)^{1+\beta_B}(\alpha_B-\beta_B)^{\alpha_B-\beta_B}
\exp(\beta_B-\alpha_B)
 & {\rm for} \ \nu'/\nu'_0 \ge \alpha_B - \beta_B \ ,
\end{array} \right.
\label{eq:spectrum}
\end{equation}
where $\nu'_0$, $\alpha_B$, and $\beta_B$ are
the break energy and the low- and high- energy photon index, respectively.
Equations (\ref{eq:jetthin}) and (\ref{eq:spectrum})
are the basic equations to calculate the flux of a single pulse.
%
%
The observed flux depends on  nine parameters:
$\gamma$, $\alpha_B$, $\beta_B$, $\Delta\theta$,
$A_0\gamma^4$, $r_0/\beta c\gamma^2$, $\gamma\nu'_0$, $z$,
and $\theta_v$.

\section{The Maximum Distance of the Observable {\it BeppoSAX}-XRFs}
\label{sec:index}
In this section,
we calculate the observed peak flux and the photon index in the energy
band 2--25\,keV as a function of
the viewing angle $\gamma\theta_v$.
The adopted parameters are $\Delta\theta=0.1$,
$\alpha_B=-1$, $\beta_B=-2.5$, $\gamma\nu'_0=300$~keV and
$r_0/\beta c\gamma^2=10$~s \citep{preece2000}.
We fix the amplitude $A_0\gamma^4$ so that
the isotropic equivalent $\gamma$-ray energy
$\Eiso=4\pi d_L^2 (1+z)^{-1}S$(20--2000~keV)
satisfies the  condition
\begin{equation}
\f{1}{2}(\Delta\theta)^2\Eiso\,=E_\gamma\ (=  {\rm const}.)\ ,
\label{eq:EgamCondition}
\end{equation}
when $\theta_v=0$.
In this section, we take the standard energy constant 
$E_\gamma=1.15\times10^{51}$~ergs
\citep{bloom2003ser}.
Then we obtain
$\gamma^4A_0=2.6\times10^8$~erg~cm$^{-2}$.
The redshift is varied from $z=0.01$ to 1.0.
 
\begin{figure}
\plotone{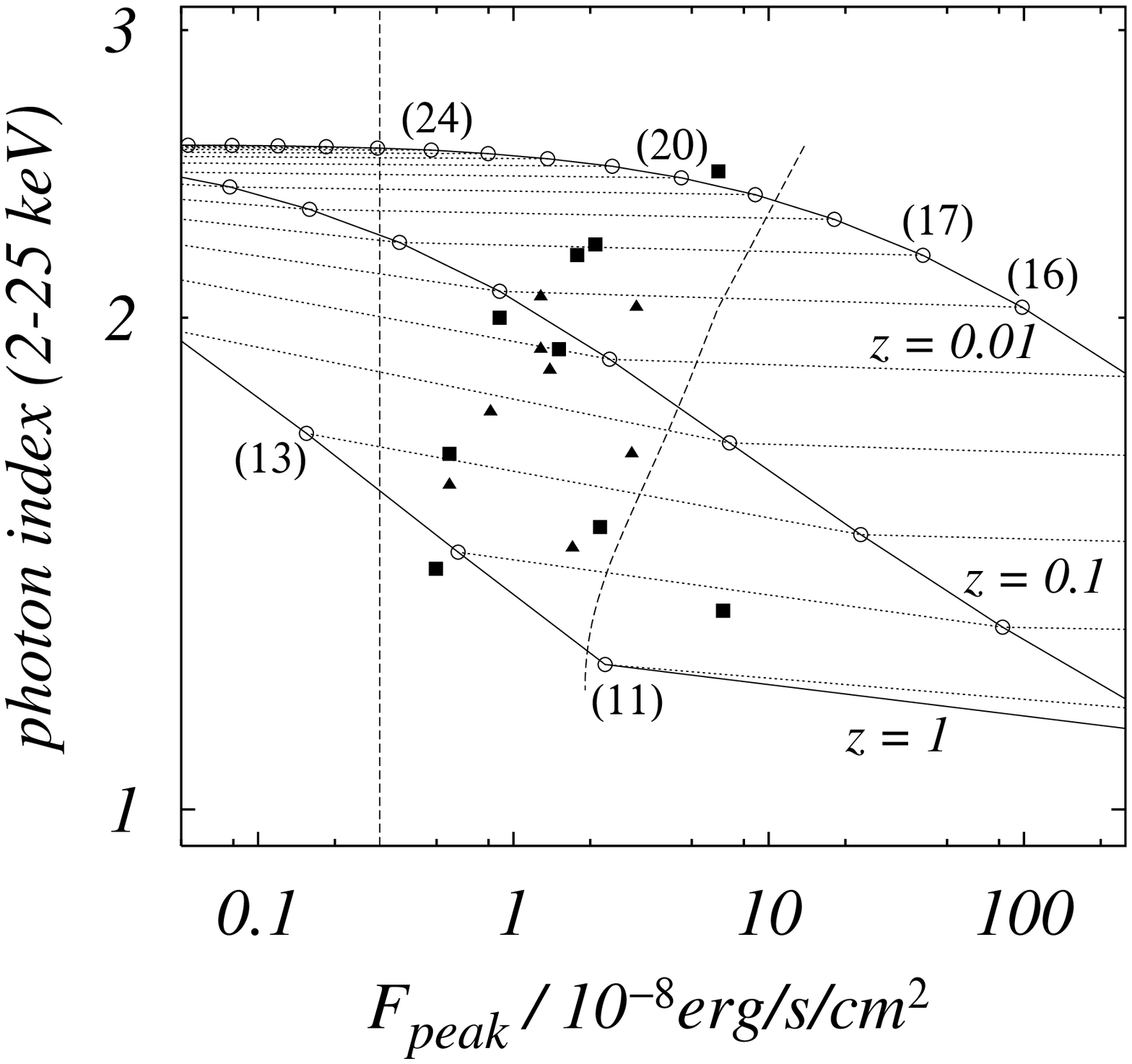}
\caption{
Photon index in the energy range 2--25~keV
as a function of the peak flux in the same energy range
by varying the source redshift $z$.
This figure is the updated version of Fig.~3 in \citet{yamazaki2002}.
We adopt $\gamma\Delta\theta=10$,
$\alpha_B=-1$, $\beta_B=-2.5$, and $\gamma\nu'_0=300$~keV.
The values of the viewing angle $\gamma\theta_v$ are
given in parenthesis.
Three solid curves correspond to
$z=0.01$, 0.1, and 1, respectively.
The same values of $\gamma\theta_v$
with different $z$ are connected by dotted lines.
The observed data of {\it BeppoSAX}-XRFs are shown from
\citet{heise2001}.
Squares (triangles) are those which were (were not)
detected  by BATSE.
Two dashed lines represent observational bounds.
Note that an operational definition of the XRF detected by
Wide Field Cameras (WFCs) on {\it BeppoSAX} is
a fast X-ray transient
that is not triggered and not detected by the Gamma-Ray Burst Monitor
(GRBM) \citep{heise2001}.
In the region to the left of the vertical dashed line,
the peak flux in the  X-ray band is smaller than the
limiting sensitivity of WFCs,
and such events cannot be observed.
In the region to the right of the oblique dashed line,
the peak flux in the $\gamma$-ray band is larger than the limiting
sensitivity of the GRBM,
and such events are observed as GRBs.
}
\label{fig_1}
\end{figure}

For our newly adopted parameters and the spectral function
in equation (\ref{eq:spectrum}),
we use a revised version of Fig.~3 in \citet{yamazaki2002},
which originally assumed a different functional form of $f(\nu')$
and used the old parameters
$E_\gamma=0.5\times10^{51}$~erg \citep{frail2001} and $\beta_B=-3$.
Figure\,\ref{fig_1} shows the results.
Although qualitative differences between old and new versions 
are small, large quantitative differences exist.
Since we now take into account the
cosmological effects that were entirely neglected in the
previous version,
the observed spectrum becomes softer at higher $z$.
The dotted lines in Figure\,\ref{fig_1} connect
the same values of $\gamma\theta_v$ with different~$z$.
 The observed XRFs take place up to
$z\sim 1$ in contrast to our previous result of $z=0.4$ and have
viewing angles $\Delta\theta\lesssim\theta_v\lesssim2\Delta\theta$.
The reason for this difference comes from the increase of the jet
energy, the different spectrum and the different high-energy photon 
index.
It is interesting to note that the  only known redshift for  
XRFs so far is $z=0.25$, one of the nearest bursts ever 
detected \citep{sakamoto2003,soderberg2003}.
 
We roughly estimate the event rate of the XRF detected by
WFC/{\it BeppoSAX} \citep{yamazaki2002}.  From the above results,
the jet emission with an opening half-angle $\Delta\theta$
is observed as the XRF (GRB) when the viewing angle
is within $\Delta\theta\lesssim\theta_v\lesssim2\Delta\theta$
($0\lesssim\theta_v\lesssim \Delta\theta$).
Therefore, the ratio of each solid angle is estimated as
$f_{\rm XRF}/f_{\rm GRB}\sim (2^2-1^2)/1^2=3$.
Using this value, we obtain
$
R_{\rm XRF}\sim 1\times10^3\,{\rm events}\ {\rm yr}^{-1}$
for the distance to the farthest XRF $d_{\rm XRF}=6$~Gpc
\citep[see equation (5) of][]{yamazaki2002}.

The derived value is comparable to the observation
or might be an overestimation that may be reduced since
the flux from the source located at $d_{\rm XRF}\sim6$~Gpc
is too low to be observed if the viewing angle $\theta_v$
is as large as $\sim2\Delta\theta$.
The ratio of the event rates of GRBs, X-ray rich GRBs (XR-GRBs)
and XRFs detected by {\it HETE-2} will be  discussed
in the following sections.

\section{$\Ep$--$\Eiso$ Relation of {\it HETE} Bursts}
\label{sec:EpEiso_XRF}
 
In this section, we perform Monte Carlo simulations in order to show that
our off-axis jet model can derive  the observed $\Ep$--$\Eiso$ relation
and the event rate of the XRFs, the XR-GRBs and
the GRBs detected by {\it HETE-2}.
We randomly generate $10^4$ bursts, each of which has
the observed flux given by equations (\ref{eq:jetthin}) and
(\ref{eq:spectrum}).
In order to calculate the observed spectrum and fluence from each
burst, we need eight parameters:
$\gamma$\,, $\alpha_B$\,, $\beta_B$\,, $\Delta\theta$,
$A_0\gamma^4(r_0/\beta c\gamma^2)^2$\,, $\gamma\nu'_0$\,, $z$, and
$\theta_v$.
They are determined in the following procedure.
\begin{enumerate}
\item We fix $\gamma=100$.
The parameters $\alpha_B$, $\beta_B$, and $\Delta\theta$
are allowed to have the following distributions.
The distribution of the low-energy (high-energy) photon index
$\alpha_B$ ($\beta_B$) is assumed
to be normal with an average of $-1$ ($-2.3$) and a standard deviation of
$0.3$ ($0.3$)
\citep{preece2000}.
The distribution of the opening half-angle
of the jet, $\Delta\theta$, is fairly unknown.
Here we assume a power-law form given as
$f_{\Delta\theta}\,d(\Delta\theta)\propto
(\Delta\theta)^{-q}\,d(\Delta\theta)$ for
$\Delta\theta_\min<\Delta\theta<\Delta\theta_\max$.
We take $q=2$ for the fiducial case,
and adopt $\Delta\theta_\max=0.3$ and
$\Delta\theta_\min=0.03$~rad, which correspond to the maximum
and minimum values inferred from observations, respectively
\citep{frail2001,panaitescu2002,bloom2003ser}.
\item
Second, we choose $\Eg\intrin$, which is
the geometrically-corrected $\gamma$-ray energy of the source in
the case of $z=0$ and $\theta_v=0$,
according to the narrow log-normal distribution
with an average and a standard deviation of
$51+\log(1.15)$ and $0.3$, respectively, for $\log (\Eg\intrin/ 1\ {\rm erg})$
\citep{bloom2003ser}.
Then, the isotropic equivalent $\gamma$-ray energy for $z=0$ and $\theta_v=0$
is calculated as $\Eiso\intrin=2(\Delta\theta)^{-2}\Eg\intrin$
to determine the flux normalization
$A_0\gamma^4(r_0/\beta c\gamma^2)^2$.
\item
Third, we assume the {\it intrinsic} $\Ep$--$\Eiso$ relation
for $z=0$ and $\theta_v=0$:
\begin{equation}
\Ep\intrin=100\,\xi\ {\rm keV}\
\left(\f{\Eiso\intrin}{10^{51}{\rm ergs}}\right)^{1/2}.
\label{eq:EpEiso_intrinsic}
\end{equation}
This may be a consequence of the standard synchrotron shock model
\citep{zhang2002,ioka2002},
but we do not discuss the origin of this intrinsic relation
in this Letter.
The coefficient $\xi$ is assumed to obey
the log-normal distribution \citep{ioka2002},
where an average and a standard deviation of $\log\xi$ are set to
$-0.7$ and $0.15$, respectively.
We determine $\gamma\nu'_0$ such that the calculated
spectrum $\nu S_\nu$ has a peak energy $\Ep\intrin$ when
$\theta_v=0$ and $z=0$.
\item Finally, we choose the source redshift $z$ and
the viewing angle $\theta_v$ to calculate the observed spectrum
and fluence, and find $\Ep$ and $\Eiso$.
The source redshift distribution is assumed to trace the cosmic star
formation rate, and the probability distribution of $\theta_v$ is
$P(\theta_v)\,d\theta_v\propto\sin\theta_v\,d\theta_v$.
To determine the redshift distribution, we assume the model SF2
of the star formation rate given by \citet{porciani2001}.
\end{enumerate}

We place a fluence truncation of
$5\times10^{-8}$~erg~cm$^{-2}$ to reflect the limiting sensitivity
of detectors on {\it HETE-2}.
Although the detection conditions of instruments vary with many
factors of each event \citep{band2002}, we consider a very simple
criterion here.
%
This fluence truncation condition is also adopted in
\citet{zhang2003st}.
 
\begin{figure}
\plotone{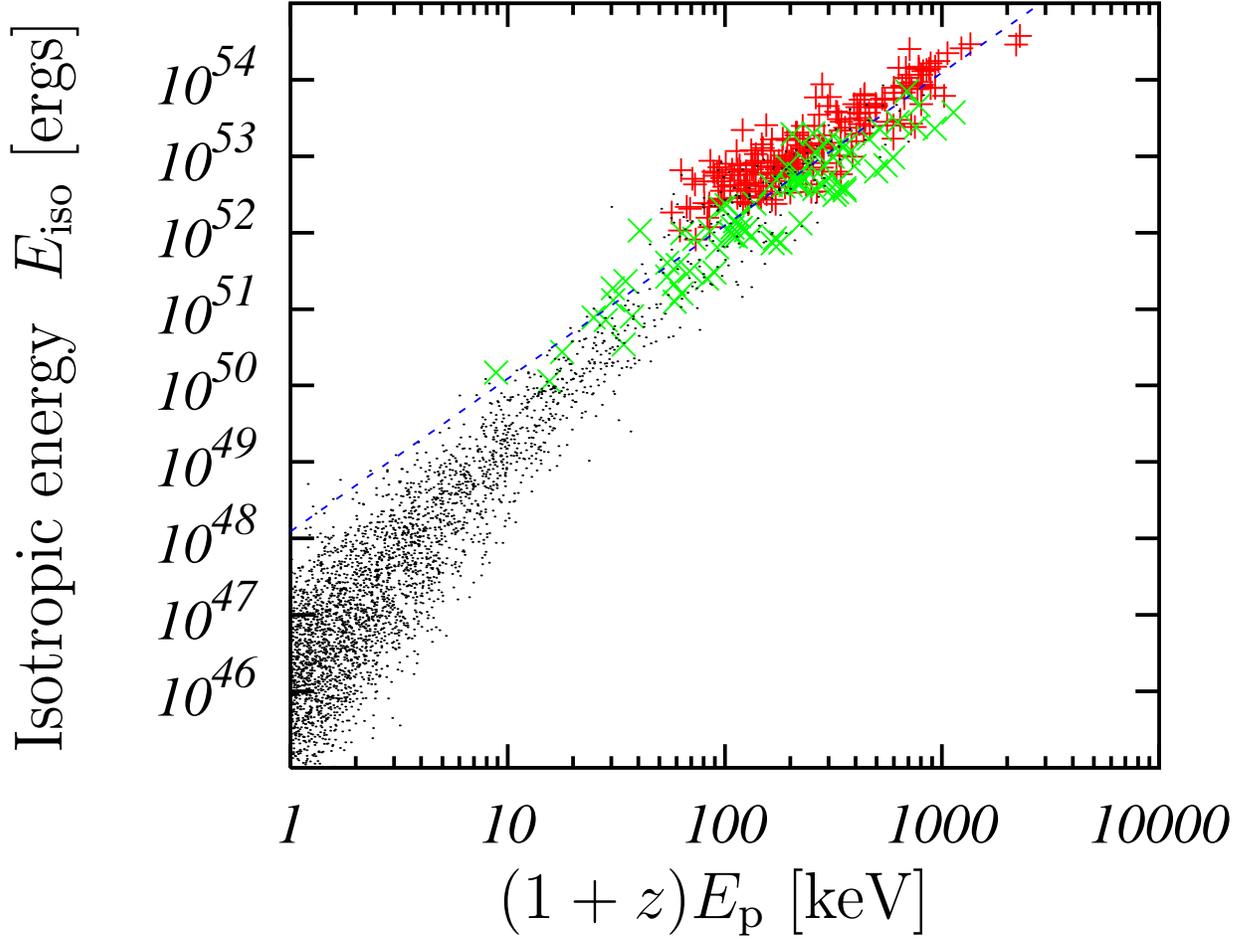}
\caption{
Distribution of simulated bursts in the $(1+z)\Ep$--$\Eiso$ plane.
Pluses ($+$) and crosses ($\times$) represent bursts that can be detected by
{\it HETE-2}; the former is on-axis events ($\theta_v<\Delta\theta$)
while the latter is the off-axis case ($\theta_v>\Delta\theta$).
The events denoted by dots are not detected.
The dashed line represents the best fit of the observation given by
$\Ep\sim95\,{\rm keV}\,(\Eiso/10^{52}\,{\rm ergs})^{1/2}$
\citep{lamb2003}.
}
\label{fig_2}
\end{figure}

Figure~\ref{fig_2} shows a result.
Among $10^4$ simulated events, 288 events are detected by {\it HETE-2}.
The others cannot be observed because their viewing angles are
so large that the relativistic beaming effect reduces their observed
flux below the limiting sensitivity.
Pluses~($+$) and crosses~($\times$) represent bursts detected by
{\it HETE-2}; the former corresponds to
 on-axis events ($\theta_v<\Delta\theta$)
while the latter correspond to off-axis events ($\theta_v>\Delta\theta$).
The events denoted by dots are not detected.
The numbers of on-axis and off-axis events are 209 and 79,
respectively.
Nearby events ($z\lesssim1$) with large viewing angles can be seen.
Such bursts are mainly  soft events with $(1+z)\Ep$
less than $\sim60$~keV.
 
When $\theta_v<\Delta\theta$, $\Ep$ is related to $\Eiso$ as
$\Ep\propto\Eiso^{1/2}$ [see equation (\ref{eq:EpEiso_intrinsic})].
The dispersion of pluses~($+$) in the $\Ep$--$\Eiso$ plane comes mainly
from those of ``intrinsic'' quantities such as
$\Eg\intrin$, $\Delta\theta$  and $\xi$.
 
On the other hand,
even when $\theta_v>\Delta\theta$, the relation
$\Ep\propto\Eiso^{1/2}$ is nearly satisfied for the observed sources.
The reason is as follows.
For a certain source, as the viewing angle increases,
the relativistic beaming and Doppler effects reduce the observed
fluence and peak energy, respectively.
When the point source approximation is appropriate for the large-$\theta_v$
case, the isotropic energy and the peak energy depend on
the Doppler factor
$\delta=[\gamma(1-\beta\cos(\theta_v-\Delta\theta))]^{-1}$ as
$\Eiso\propto S(20-2000~{\rm keV})\propto\delta^{1-\malpha}$ and
$\Ep\propto\delta$, respectively
\citep{ioka2001,yamazaki2002,dar2003}.
Hence we obtain $\Eiso\propto\Ep^{1-\malpha}$.
Here $\malpha$ is the mean photon index in the 20--2000~keV band, which
ranges between $\beta_B$ and $\alpha_B$.
Therefore we can explain the relation $\Ep\propto\Eiso^{1/2}$
for $\malpha\sim\alpha_B \sim -1$.
On the other hand, when $\theta_v$ is large enough for $\Ep$
to be smaller than 20~keV,
we find $\Eiso\propto\Ep^{1-\beta_B}\sim\Ep^{3.3}$ or $\Ep \propto \Eiso^{0.3}$
since $\malpha\sim\beta_B$.
In this case, the relation deviates from the line
$\Ep\propto\Eiso^{1/2}$, and the
dispersion of $\Eiso$ becomes large for small $\Ep$.

\section{Discussion}
\label{sec:dis}
 
%
We have shown that our simple jet model does not contradict the
observed $\Ep$--$\Eiso$ relation and extends it to lower
$\Ep$ or $\Eiso$ values.
The low-isotropic energy part of the relation is dominated by
off-axis events.
The number of off-axis events is about one-third of on-axis emissions.
An important prediction of our model has been also derived, i.e.,
we will see the deviation from the present relation
$\Ep\propto\Eiso^{1/2}$ if the statistics of the low-energy bursts
increase.
%

{\it HETE} team gives  definitions of the XRF and
the XR-GRB in terms of the hardness ratio:
XRFs and XR-GRBs are  events for which
$\log[S_X(2-30~{\rm keV})/S_\gamma(30-400~{\rm keV})]>0.0$ and
$-0.5$, respectively \citep{lamb2003,sakamoto2003}.
We calculate  the hardness ratio for simulated bursts surviving the
fluence truncation condition, and classify them into
GRBs, XR-GRBs, and XRFs.
It is then found  that all XRFs have redshift smaller than 5.
The ratio of the observed event rate becomes
$R_{GRB}:R_{XR-GRB}:R_{XRF}\sim2:6:1$.
This ratio mainly depends on the value of $q$.
When $q$ becomes small, jets with large $\Delta\theta$ increases,
and hence intrinsically dim bursts (i.e., low-$\Eiso\intrin$ bursts)
are enhanced.
Owing to equation (\ref{eq:EpEiso_intrinsic}), 
soft events are enhanced.
In the case of $q=1$ with the other parameters remaining
fiducial values,
the ratio is $R_{GRB}:R_{XR-GRB}:R_{XRF}\sim1:9:3$.
For any cases we have done, the number of XR-GRBs is larger than
those of GRBs and XRFs, while the event rate is essentially
comparable with each other.
{\it HETE-2} observation shows $R_{GRB}:R_{XR-GRB}:R_{XRF}\sim1:1:1$
\citep{lamb2003}.
Although possible instrumental biases may change the observed ratio
(Suzuki,~M. \& Kawai,~N., 2003, private communication),
we need more studies in order to bridge a small gap between
the theoretical and the observational results.

We briefly comment on how the results obtained in this Letter will
depend on the Lorentz factor of the jet $\gamma$.
If we fix $\gamma=200$, the relativistic beaming effect becomes 
stronger and less off-axis events are observed than in the case of
$\gamma=100$;
off-axis events are 13~\% of the whole observed bursts when
$\gamma=200$, while 27~\% for $\gamma=100$.
The ratio of the observed event rate for $\gamma=200$ is
$R_{GRB}:R_{XR-GRB}:R_{XRF}\sim2:5:1$,
which is similar to that for $\gamma=100$.

The $\Ep$--$\Eiso$ diagram of the GRB population may be a counterpart
of the Herzsprung-Russell diagram of the stellar evolution.
The main-sequence stars cluster around a single curve which is
a one-parameter family of the stellar mass.
This suggests that the $\Ep$--$\Eiso$ relation of the GRB implies
the existence of a certain parameter that controls the GRB nature
like the stellar mass.
We have shown that the viewing angle is one main factor
to explain the $\Ep$--$\Eiso$ relation kinematically.
Our model predicts the deviation of this relation in the small $\Eiso$
region, which may be confirmed in future.
 
In the uniform jet model, 
the afterglows of off-axis jets may resemble
the orphan afterglows that initially have a rising light curve
\citep[e.g.,][]{yamazaki2003df,granot2002,tp02}.
The observed $R$-band light curve of the afterglow of XRF~030723
may support our model \citep{fynbo04}.

 
\acknowledgments
We would like to thank the referee for useful comments and
suggestions.
We would like to thank G.~R.~Ricker, T.~Murakami, N.~Kawai,
A.~Yoshida, and M.~Suzuki for useful comments and discussions.
Numerical computation in this work was carried out at the
Yukawa Institute Computer Facility.
This work was supported in part by
a Grant-in-Aid for for the 21st Century COE
``Center for Diversity and Universality in Physics''
and also supported by Grant-in-Aid for Scientific Research
of the Japanese Ministry of Education, Culture, Sports, Science
and Technology, No.05008 (RY),
No.660 (KI),
No.14047212 (TN), and No.14204024 (TN).



\clearpage

\end{document}